\documentclass[english,aps,prb,showpacs,superscriptaddress,floats,amsmath,amssymb,floatfix,nobalancelastpage,twocolumn,groupedaddress]{revtex4-2}
\usepackage[T1]{fontenc}
\usepackage[latin9]{inputenc}
\usepackage{units}
\usepackage{amsmath}
\usepackage{amssymb}
\usepackage{graphicx}

\makeatletter

\newcommand*\LyXThinSpace{\,\hspace{0pt}}

\usepackage{accents}

\makeatother

\usepackage{babel}
\graphicspath{{./}{./fig/}{./figs/}{./psfigs/}{./Figures/}}
\begin{document}
\title{Two-dimensional nonlinear optical response of a spiral magnet}
\author{Wolfram Brenig}
\email{w.brenig@tu-braunschweig.de}

\affiliation{Institute for Theoretical Physics, Technical University Braunschweig,
D-38106 Braunschweig, Germany}
\begin{abstract}
We study the dynamical response function relevant for two-dimensional
coherent nonlinear optical spectroscopy of the antiferromagnetic frustrated
$J_{1}$-$J_{3}$ Heisenberg model on the square lattice within its
long-range ordered, incommensurate diagonal spiral phase.
We argue that in this phase effective dipole coupling to the electric
field is important, with the spin-current coupling potentially being
the dominant mechanism for spin-$\nicefrac{1}{2}$. For this setting,
we use linear spin wave theory to evaluate the leading nonlinear polarization
response which is of second order in the driving field. We show that
the response function features a strong antidiagonal, galvanoelectric
feature. The width of this feature is set by relaxation rates beyond
the noninteracting magnon picture, thereby providing access to single-magnon
lifetimes within the multi-magnon continuum of the response function.
Moreover, the response function is shown to display various structures
in the two-dimensional frequency plane related to exceptional regions
of the magnon dispersion.
\end{abstract}
\maketitle

\section{Introduction}

Two-dimensional (2D) nonlinear coherent spectroscopy (CS) \citep{Mukamel2000,Cho2008,Cundiff2013}
has recently experienced an upsurge of interest as a dynamical probe
of low-dimensional quantum magnets. This has been triggered in particular
by the seminal suggestion that properties of single quasiparticles,
which result from spin-fractionalization should be accessible by 2DCS
\citep{Wan2019}. This approach is different from the standard analysis
of fractionalized spin dynamics by inelastic neutron scattering, where
ubiquitous multiparticle continua mask the single particle properties
\citep{Stone2003,Lake2005}. In turn a significant number of theoretical
studies of nonlinear optical spectroscopy in quantum magnets, prone
to fractionalization, has emerged. This pertains to spinons in (quasi-)1D
spin-chain systems \citep{Li2021,Sim2023L,Gao2023,Sim2023,ZLLi2023,Potts2023,Watanabe2024,Srivastava2025},
as well as Majorana fermions and visons in Kitaev magnets \citep{Choi2020,Negahdari2023,Kanega2021,Krupnitska2023,Brenig2024,Qiang2023,Kaib2024}
and fractons in X-cube and Haah code models \citep{Nandkishore2021}.

In contrast to spin-systems with a sought-for fractionalization, quantum
magnets with long-range order (LRO) resulting from broken spin-rotational
invariance and exhibiting magnon excitations have received much less
attention regarding 2DCS. Existing early analysis of canted antiferromagnets
\citep{Lu2017} has focused on 2D nonlinear electron spin resonance,
driven by Zeeman coupling to external \emph{magnetic} fields. In contrast
to this, coupling magnets to \emph{electric} fields, i.e., by the
magnetoelectric effect (MEE) \citep{Dzyaloshinskii1960}, is another
important option \citep{Fiebig2005,Tokura2014,Dong2015,Dong2016}.
While the MEE is of prime interest in the context of static phenomena,
electromagon excitations \citep{Baryakhtar1970,Smolenskii1982} show
that dynamics is of equal importance \citep{Takahashi2012} and may
lead to observable consequences in optical probing \citep{Azimi2016,Zhang2021,Ueda2022}.

In turn, the motivation for this work is to pursue a scenario where
the 2D nonlinear response of a quantum magnet is driven primarily
by coupling to electric fields through the MEE. Moreover, we seek
for a situation in which only a\emph{ }single microscopic mechanism
is relevant for the MEE. For that, we lay out the favorable conditions
as follows: First, we require magnetic LRO. Then, the elementary excitations
are magnons, which implies that any nonlinear dynamical response,
driven by Zeeman coupling of the magnetization to magnetic fields,
is necessarily an effect involving magnon-magnon interactions. We
assume them to be of subleading order and therefore drop coupling
to magnetic fields. Second, we consider the most prominent microscopic
mechanisms for MEE coupling more closely \citep{Tokura2014}. To begin,
there is the spin-dependent hybridization \citep{Jia2007}. In this
work, we will assume a spin-$\nicefrac{1}{2}$ system given, for which
this mechanism is strictly zero \citep{ArbS}. Next, we focus on the
exchange-striction coupling \citep{Katsura2009}. This requires some
sort of inversion symmetry breaking and is a relevant option for quasi
one-dimensional systems. In this work however, we will remain with
planar square lattices, for which inversion symmetry breaking is less
likely. In turn we also dismiss exchange-striction coupling. Somewhat
related to exchange-striction are dynamic couplings to electric fields
involving phonon assistance \citep{Lorenzana1995}. These require
favorable optical vibrational modes which we waive for the present
work. This leaves the spin-current coupling or inverse Dzyaloshinskii--Moriya
(DM) interaction, introduced by Katsura, Nagaosa, and Balatsky (KNB)
\citep{Katsura2005}. Quite generally, the KNB mechanism allows for
a MEE, both in quasi-1D situations with chiral or spiral correlations,
e.g., \citep{Richter2022} and refs. therein, as well as in systems
with non-collinear LRO \citep{Sergienko2006,Cheong2007} in higher
dimension. While for any specific material realization a non-zero
coupling constant for the KNB mechanism remains a question to be clarified,
we focus on this form of light-matter coupling.

Summing up, we will consider a 2D quantum magnet, which allows for
spiral LRO. For that purpose we select the frustrated $J_{1}$-$J_{3}$
antiferromagnetic (AFM) Heisenberg model on the square lattice (HSL).
We subject this to a time-dependent electric field which couples to
the spin system via the KNB mechanism. For this setting, we will evaluate
the leading order 2D nonlinear response function (NRF). We will analyze
the spectral properties of the 2D NRF and show in particular that
it displays a giant galvanoelectric effect, that it allows to extract
magnon lifetimes, and that it displays specific signatures of the
magnon DOS. The paper is organized as follows: In Sec. \ref{sec:Model},
we detail the model and its excitations as well as the light-matter
coupling. Sec. \ref{sec:2DCS-Response-Fun} describes the calculation
of the 2D NRF. In Sec. \ref{sec:Discussion}, we discuss the central
features of the 2D NRF. We summarize in Sec. \ref{sec:Summary}. Appendix
\ref{sec:comm} sketches an alternative calculational approach.

\section{Model\label{sec:Model}}

In this section, we first detail the linear spin wave theory (LSWT)
of the $J_{1}$-$J_{3}$ HSL. Second, we explain the KNB-polarization.

\subsection{Spiral phase of the $J_{1}$-$J_{3}$ antiferromagnet}

\begin{figure}[tb]
\centering{}\includegraphics[width=0.6\columnwidth]{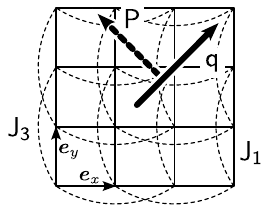}\caption{Frustrated AFM $J_{1}$-$J_{3}$ model on square lattice including
ICS pitch vector $\boldsymbol{q}$ and classical KNB polarization
$\boldsymbol{P}$, Eq. (\ref{eq:Pcl}). \label{fig:model}}
\end{figure}

Before starting the theoretical developments, we emphasize that our
aim is not to add to the large body of work on the phases of $J_{1}$(-$J_{2})$-$J_{3}$
quantum antiferromagnets on the square lattice. This model has come
under scrutiny early on, in the context of the cuprate superconductors
\citep{Gelfand1989,Figueirido1990,Moreo1990,Chubukov1991} and remains
of great interest until today, see \citep{Liu2024} and refs. therein.
Here, we rather use the $J_{1}$-$J_{3}$ model as an established
device, which allows to safely claim the existence of a parameter
region hosting an incommensurate spiral (ICS) state with LRO \citep{NoJ2}. With
that in mind, we describe our LSWT of that ICS state. The Hamiltonian
reads
\begin{equation}
H/J_{1}=\sum_{\left\langle \boldsymbol{l}\boldsymbol{m}\right\rangle }\boldsymbol{S}_{\boldsymbol{l}}\cdot\boldsymbol{S}_{\boldsymbol{m}}+j\sum_{\left\langle \left\langle \left\langle \boldsymbol{l}\boldsymbol{m}\right\rangle \right\rangle \right\rangle }\boldsymbol{S}_{\boldsymbol{l}}\cdot\boldsymbol{S}_{\boldsymbol{m}}\,,\label{eq:h}
\end{equation}
where $\boldsymbol{l}=\boldsymbol{e}_{x}l_{x}+\boldsymbol{e}_{y}l_{y}$,
and we set the lattice constant $a\equiv1$, i.e., $\boldsymbol{e}_{x,y}=(1,0),\,(0,1)$,
see Fig. \ref{fig:model}. $\boldsymbol{S}_{\boldsymbol{l}}$ are
spin operators with $\boldsymbol{S}_{\boldsymbol{l}}^{2}=S(S+1)$.
As usual, $S$ will be kept as a free parameter of the LSWT, although
we have $S=1/2$ in mind. We normalize all energies with respect to
$J_{1}$, i.e., $j=J_{3}/J_{1}$, where $J_{1}$($J_{3}$) are nearest(third-nearest)-neighbor
exchange couplings. As for the classical phases of this model \citep{Gelfand1989,Moreo1990}
and staying with $j\geq0$, for $0\leq j<1/4\equiv j_{c}$, the ground
state is a standard nearest-neighbor (NN) AFM. For $j_{c}<j$, the
system acquires an ICS with a four-fold degenerate 2D pitch angle
of $\boldsymbol{q}=(\pm q,\pm q)$ where $q=\arccos(1/(4j))$. While
the analysis of the quantum phases of the more general $J_{1}$-$J_{2}$-$J_{3}$
model including also a next-nearest-neighbor exchange $J_{2}$ has
a long standing history, the most recent status of this establishes
that the ICS remains a robust feature of the model \citep{Liu2024},
even if quantum fluctuations shift the classical transition point
substantially from $j_{c}=1/4$ to larger values on the order of $j_{c}\sim1$.
Based on this, we will accept that ICS order is a valid assumption
for the $J_{1}$-$J_{3}$ model above a critical value of $j$, and
use linear spin wave theory (LSWT) to treat the elementary excitations
in that regime - even if the value of $j_{c}$ lacks quantum corrections
and therefore is too small in LSWT.

\begin{figure}[tb]
\centering{}\includegraphics[width=0.45\columnwidth]{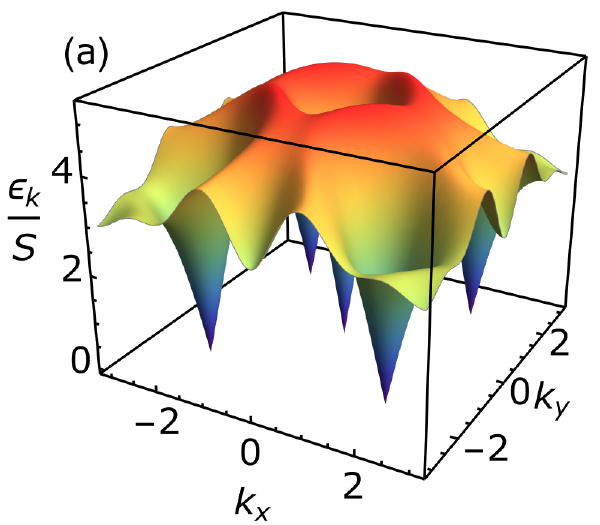}\hskip .3cm\includegraphics[width=0.47\columnwidth]{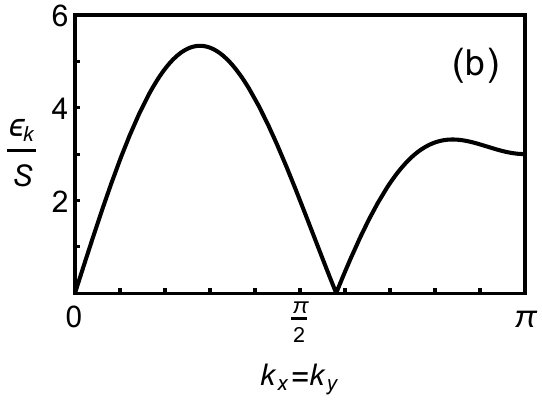}\caption{LSWT magnon dispersion for $j=J_{3}/J_{1}=1$, normalized to $J_{1}$.
(a) 3D view, (b) 2D cut along the diagonal $k_{x}\allowbreak=\allowbreak k_{y}$.\label{fig:dispplt}}
\end{figure}

For all calculations, we choose one out of of the four degenerate
pitch vectors, i.e., $\boldsymbol{q}=(q,q)$, see Fig. \ref{fig:model},
and rotate onto a locally ferromagnetic (FM) coordinate frame with
spins $\tilde{\boldsymbol{S}}_{\boldsymbol{l}}$ 
\begin{equation}
\boldsymbol{S}_{\boldsymbol{l}}=\left[\begin{array}{ccc}
0 & -\sin(\boldsymbol{q}\cdot\boldsymbol{l}) & \cos(\boldsymbol{q}\cdot\boldsymbol{l})\\
0 & \cos(\boldsymbol{q}\cdot\boldsymbol{l}) & \sin(\boldsymbol{q}\cdot\boldsymbol{l})\\
-1 & 0 & 0
\end{array}\right]\tilde{\boldsymbol{S}_{\boldsymbol{l}}}\,.\label{eq:RS-1}
\end{equation}
The ICS is obvious, inserting a classical FM state $\tilde{\boldsymbol{S}_{\boldsymbol{l}}}=(0,0,1)\,S$.
Describing quantum fluctuations off this state by Holstein-Primakoff
(HP) bosons requires only a \emph{single site} magnetic unit cell,
with
\begin{align}
\tilde{S}_{\boldsymbol{l}}^{z}= & S-a_{\boldsymbol{l}}^{\dagger}a_{\boldsymbol{l}}\,,\nonumber \\
\tilde{S}_{\boldsymbol{l}}^{+}= & (2S-a_{\boldsymbol{l}}^{\dagger}a_{\boldsymbol{l}})^{1/2}a_{\boldsymbol{l}}\,,\nonumber \\
\tilde{S}_{\boldsymbol{l}}^{-}= & a_{\boldsymbol{l}}^{\dagger}(2S-a_{\boldsymbol{l}}^{\dagger}a_{\boldsymbol{l}})^{1/2}\,.\label{eq:HPB-1}
\end{align}
Performing the usual expansion of the Hamiltonian to leading $O(1/S)$
in terms of these HP boson, i.e., linear spin wave theory (LSWT),
we arrive at
\begin{equation}
H=\sum_{\boldsymbol{k}}\boldsymbol{A}_{\boldsymbol{k}}^{+}\boldsymbol{h}_{\boldsymbol{k}}\boldsymbol{A}_{\boldsymbol{k}}^{\phantom{+}}+NS(S+1)e_{cl}\,,
\end{equation}
where $N$ is the number of sites and $e_{cl}=-(2j+1/(4j))$ is the
classical energy per site for $S=1$. $\boldsymbol{A}_{\boldsymbol{k}}^{+}=(a_{\boldsymbol{k}}^{\dagger},a_{-\boldsymbol{k}})$
is a boson spinor and
\begin{align}
\boldsymbol{h}_{\boldsymbol{k}}= & \frac{S}{2}\begin{bmatrix}\mathcal{A}_{\boldsymbol{k}} & \mathcal{B}_{\boldsymbol{k}}\\
\mathcal{B}_{\boldsymbol{k}} & \mathcal{A}_{\boldsymbol{k}}
\end{bmatrix}\,,\\
\mathcal{A}_{\boldsymbol{k}}= & \smash[b]{\frac{S}{8j}}(4+32j^{2}+(8j-2)(\cos(k_{x})+\nonumber \\
 & \hphantom{\smash[b]{\frac{S}{16j}a}}\cos(k_{y}))+\cos(2k_{x})+\cos(2k_{y}))\label{Ak}\\
\mathcal{B}_{\boldsymbol{k}}= & \smash[b]{\frac{S}{8j}}(4j+1)((4j-1)(\cos(2k_{x})+\cos(2k_{y}))+\nonumber \\
 & \hphantom{\smash[b]{\frac{S}{16j}}a}2\cos(k_{x})+2\cos(k_{y}))
\end{align}
All bold faced spinor operators in this work will also be referenced
by their components, using the notation $A_{k\mu}^{(+)}$, with non-bold
letters and subscripts $\mu=1,2.$ The Hamiltonian can be diagonalized
by means of a Bogoliubov transformation $\boldsymbol{A}_{\boldsymbol{k}}=\boldsymbol{U}_{\boldsymbol{k}}\boldsymbol{D}_{\boldsymbol{k}}$,
onto diagonal bosons $\boldsymbol{D}_{\boldsymbol{k}}^{+}=(d_{\boldsymbol{k}}^{\dagger},d_{-\boldsymbol{k}})$
which create and destroy magnon excitations
\begin{equation}
H=\smash[b]{\sum_{\boldsymbol{k}}}\epsilon_{\boldsymbol{k}}d_{\boldsymbol{k}}^{\dagger}d_{\boldsymbol{k}}^{\phantom{\dagger}}+E_{0}\,,\label{eq:digh}
\end{equation}
where
\begin{equation}
\boldsymbol{U}_{\boldsymbol{k}}=\begin{bmatrix}u_{\boldsymbol{k}} & v_{\boldsymbol{k}}\\
v_{\boldsymbol{k}} & u_{\boldsymbol{k}}
\end{bmatrix}\,,\hphantom{aa}\boldsymbol{U}_{\boldsymbol{k}}^{T}\boldsymbol{h}_{\boldsymbol{k}}\boldsymbol{U}_{\boldsymbol{k}}=\frac{1}{2}\begin{bmatrix}\epsilon_{\boldsymbol{k}} & 0\\
0 & \epsilon_{\boldsymbol{k}}
\end{bmatrix}\,,\label{eq:bogo1}
\end{equation}
with
\begin{equation}
u_{\boldsymbol{k}}=\sqrt{\frac{\mathcal{A}_{\boldsymbol{k}}+\epsilon_{\boldsymbol{k}}}{2\epsilon_{\boldsymbol{k}}}}\,,\hphantom{aa}v_{\boldsymbol{k}}={-}\text{sgn}(\mathcal{B}_{\boldsymbol{k}})\sqrt{\frac{\mathcal{A}_{\boldsymbol{k}}-\epsilon_{\boldsymbol{k}}}{2\epsilon_{\boldsymbol{k}}}}\,.\label{eq:bogo2}
\end{equation}
where $\epsilon_{\boldsymbol{k}}$ is the magnon energy
\begin{align}
\epsilon_{\boldsymbol{k}}= & \sqrt{\mathcal{A}_{\boldsymbol{k}}^{2}-\mathcal{B}_{\boldsymbol{k}}^{2}}\\
= & \smash[b]{\frac{S}{8j}}[(4+32j^{2}+(8j-2)(\cos(k_{x})+\cos(k_{y}))+\nonumber \\
 & \hphantom{\frac{S}{8j}[}\cos(2k_{x})+\cos(2k_{y}))^{2}-((16j^{2}-1)\times\nonumber \\
 & \hphantom{\frac{S}{8j}[}(\cos(2k_{x})+\cos(2k_{y}))+2(4j+1)\times\nonumber \\
 & \hphantom{\frac{S}{8j}[}(\cos(k_{x})+\cos(k_{y})))^{2}]^{1/2}\,,\label{ek}
\end{align}
and $E_{0}=NS(S+1)e_{cl}+\sum_{\boldsymbol{k}}\epsilon_{\boldsymbol{k}}/2$, where
the second term is the usual zero-point energy of the harmonic Bose
gas encoded in Eq. (\ref{eq:digh}). Fig. \ref{fig:dispplt} displays
the dispersion for a typical value of $j=1$, which we will remain
with for this work. The diagonal cut in this figure clearly demonstrates
the incommensurate pitch which is at $q\simeq0.58\,\pi$.

Apart from $j=1$, we mention two special cases explicitly. The first
is $j\rightarrow\infty$, where we remain with $\epsilon_{\boldsymbol{k}}=4jS[1-(\cos(2k_{x})+\cos(2k_{y}))^{2}\allowbreak/4]^{1/2}$.
This is exactly the dispersion of AFM magnons by LSWT on the decoupled
square lattices with unit cell $2a=2$ which stays effective in that
limit. The presence of four interpenetrating of these lattices is
taken into account by the Brillouin zone (BZ) still extending over
$k_{x,y}=[-\pi,\pi]$. The second case is $j=j_{c}=1/4$, where $\epsilon_{\boldsymbol{k}}=S((\allowbreak\cos(2k_{x})+\allowbreak\cos(2k_{y})+\allowbreak6)^{2}-\allowbreak16(\cos(k_{x})+\cos(k_{y}))^{2})^{1/2}/2$.
While this does already have its zeros only at $\boldsymbol{k}_{0}=(0,0)$
and $(\pm\pi,\pm\pi)$, as in the square lattice AFM state for $j<j_{c}$,
the dispersion at these wavevectors directly at the critical point
is anisotropically quadratic with $\epsilon_{\boldsymbol{k}_{0}+\boldsymbol{p}}\simeq\sqrt{2}S(p_{x}^{4}+p_{y}^{4})^{1/2}$,
for $|\boldsymbol{p}|\ll\pi$. An analysis of the critical region
is beyond the scope of this work.

\subsection{Minimal coupling from KNB polarization}

For the $J_{1}$-$J_{3}$ model, the spin-current induced, or KNB
coupling \citep{Katsura2005} to an external dynamic electric field
$\boldsymbol{E}(t)$ reads
\begin{align}
\lefteqn{H_{\mathrm{KNB}}(t)=-\boldsymbol{P}\cdot\boldsymbol{E}(t)}\label{eq:knb}\\
 & \boldsymbol{P}/\gamma=\sum_{\left\langle \boldsymbol{l}\boldsymbol{m}\right\rangle }\boldsymbol{R}_{\boldsymbol{lm}}{\times}\boldsymbol{S}_{\boldsymbol{l}}{\times}\boldsymbol{S}_{\boldsymbol{m}}+g\!\!\!\sum_{\left\langle \left\langle \left\langle \boldsymbol{l}\boldsymbol{m}\right\rangle \right\rangle \right\rangle }\!\!\!\boldsymbol{R}_{\boldsymbol{lm}}{\times}\boldsymbol{S}_{\boldsymbol{l}}{\times}\boldsymbol{S}_{\boldsymbol{m}},\hphantom{aa}\nonumber 
\end{align}
where $\boldsymbol{P}$ is the electric polarization $\gamma$ and
$\gamma g$ are effective nearest and next-next-nearest neighbor coupling
constants. For the remainder of this work, and similar to the unit
of energy $J_{1}$, we normalize $\boldsymbol{P}$ to $\gamma$ to
abbreviate the notation. $\boldsymbol{R_{lm}}$ are nearest and next-next-nearest
neighbor lattice vectors $(1,0),\,\allowbreak(0,1)$ and $(2,0),\,\allowbreak(0,2)$.

The classical KNB polarization $\boldsymbol{P}_{cl}$ results from
inserting the classical ICS into Eq. (\ref{eq:knb})
\begin{equation}
\boldsymbol{P}_{cl}=S^{2}\frac{(g-j)\sqrt{16j^{2}-1}}{4j^{2}}(-1,1,0)\,.\label{eq:Pcl}
\end{equation}
I.e., the classical polarization lies within the lattice plane and
is perpendicular to the pitch, as in Fig. \ref{fig:model}. For any
material realization of a 2D $J_{1}$-$J_{3}$ model, there is no
a priori quantum chemical reason for the dimensionful(less) parameters
$J_{1}$, $\gamma$ ($j$, $g$) from Eq. (\ref{eq:h}) and from $\boldsymbol{P}$
to be ``identical''. Yet, it seems reasonable for them to vary similarly
with other parameters of the material, and also for $j$, $g$ to
be of similar magnitude. I.e., we will assume $j\sim g$, but not
equal. In the commensurate AFM for $j<j_{c}$, the polarization is
zero. Also for $j\rightarrow j_{c}+0$ from within the ICS, the polarization
turns to zero. For $j\rightarrow\infty$ and $g=cj$ with $c\neq1$
set to some arbitrary constant, $\boldsymbol{P}_{cl}$ from Eq (\ref{eq:Pcl})
approaches a constant. In terms of the absolute scales $\gamma$ and
$\gamma g$ of Eq. (\ref{eq:knb}) this however implies that also
in this particular limit, $\boldsymbol{P}_{cl}$ is asymptotically
negligible, if considered on the scale of $g\gamma$.

Next we express the quantum corrections to $\boldsymbol{P}_{cl}$
in terms of HP bosons. For the remainder of this work, we focus on
the components in the plane of the classical polarization, i.e., $(P_{x},P_{y})$.
To begin with, we note that in LSWT, and apart from the classical
contribution, $\boldsymbol{P}$ will not start at quadratic order
in the bosons, but rather contain linear terms also. Physically, these
refer to a direct mixing between photons and magnons. Now, for all
purposes of our analysis the wave vector of the light $\boldsymbol{q}_{E}$
satisfies $\boldsymbol{q}_{E}=\boldsymbol{0}$. This implies that
all processes, possibly related to such linear mixing, need to occur
at energies of $\epsilon_{\boldsymbol{k}=\boldsymbol{q}_{E}}=0$.
Since we are only interested in finite frequencies of the external
electric field, we discard linear boson contributions to $\boldsymbol{P}$.

Inserting the HP bosons into Eq. (\ref{eq:knb}) and rotating into
the locally FM frame is lengthy, but straightforward. At quadratic
order in $\boldsymbol{A}_{\boldsymbol{k}}^{(+)}$ we obtain
\begin{align}
P_{x}= & \sum_{\boldsymbol{k}}\boldsymbol{A}_{\boldsymbol{k}}^{+}\boldsymbol{p}_{x,\boldsymbol{k}}\boldsymbol{A}_{\boldsymbol{k}}^{\phantom{+}}\\
\boldsymbol{p}_{x,\boldsymbol{k}}= & -\frac{S}{2}\frac{\sqrt{16j^{2}-1}}{4j^{2}}\left(2(j-g)\left[\begin{array}{cc}
1 & 0\\
0 & 1
\end{array}\right]+\right.\nonumber \\
 & \left.(j\cos(k_{y})-g\cos(2k_{y}))\left[\begin{array}{cc}
-1 & 1\\
1 & -1
\end{array}\right]\right)\,.\label{eq:px}
\end{align}
For $P_{y}$, an identical expression is obtained by substituting
$\boldsymbol{p}_{y,\boldsymbol{k}}=-\boldsymbol{p}_{x,\boldsymbol{k}}|_{k_{y}\rightarrow k_{x}}$.

We emphasize that Eq. (\ref{eq:px}) is obtained from an expression
which is valid at \emph{any} pitch $q$ after inserting its classical
value in the ICS for $j\geq j_{c}$. Considering the same expression
in a state of plain AFM order, i.e., at $q=\pi$ for $j<j_{c}$, simply
yields $P_{x}=0$. Speaking differently, electric field induced bilinear
coupling to magnons occurs only in the ICS.

\section{Second order two-dimensional response function\label{sec:2DCS-Response-Fun}}

Formally, inversion symmetry forces $O(2)$ NRFs to be zero. However,
the ICS breaks this symmetry. In turn, the leading order 2D response
of the polarization $\langle\Delta P^{\parallel}\rangle(t)$, projected
onto $\boldsymbol{E}_{ac}(t)$, is $O(2)$. It is described by the
Fourier transform into the 2D frequency plane of the retarded response
function $\tilde{\chi}_{2}(t,t_{1},t_{2})=i^{2}\Theta(t-t_{1})\Theta(t_{1}-t_{2})\allowbreak\langle[[P^{\parallel}(t),\allowbreak P^{\parallel}(t_{1})],P^{\parallel}(t_{2})]\rangle$
\citep{Butcher1990}. The $N$-fold time integrations in the perturbative
expansion of $\langle\Delta P^{\parallel}\rangle(t)$ at any order
$N$ in $\boldsymbol{P}\cdot\boldsymbol{E}_{ac}(t)\equiv P^{\parallel}E_{ac}(t)$
are totally symmetric with respect to any permutation of the $N$
time arguments \citep{Butcher1990}. This is termed intrinsic permutation
symmetry. In turn, any $N$-th order contributions to $\langle\Delta P^{\parallel}\rangle(t)$
requires only the fully symmetrized response function $\chi_{N}(t,t_{1},\allowbreak\dots,t_{n})=\allowbreak\sum_{M}\tilde{\chi}_{N}(t,t_{M(1)},\allowbreak\dots,t_{M(N)})/N!$
to be evaluated, where $M$ labels all permutations. The Fourier transform
of $\chi_{N}(t,t_{1},\allowbreak\dots,t_{n})$ can be obtained from
analytic continuation to the real axis of the Matsubara frequency
transform of the fully connected contractions of the imaginary time
propagator $\chi_{n}(\tau_{n},\dots\tau_{1})=\langle T_{\tau}(\allowbreak P^{\parallel}(\tau_{n})\allowbreak\dots P^{\parallel}(\tau_{1})P^{\parallel})\rangle$
\citep{Evans1966,Rostami2021}.

Diagrammatic calculations are performed in the diagonal HP bosons
basis. This involves calculation of contractions from operator groupings
of type
\begin{align}
\sum_{\dots\boldsymbol{k}_{l}\dots\boldsymbol{k}_{m}\dots} & \langle T_{\tau}(\dots\boldsymbol{D}_{\boldsymbol{k}_{l}}^{+}(\tau_{l})\boldsymbol{s}_{\boldsymbol{k}_{l}}\boldsymbol{D}_{\boldsymbol{k}_{l}}(\tau_{l})\dots\nonumber \\
 & \dots\boldsymbol{D}_{\boldsymbol{k}_{m}}^{+}(\tau_{m})\boldsymbol{s}_{\boldsymbol{k}_{m}}\boldsymbol{D}_{\boldsymbol{k}_{m}}(\tau_{m})\dots\rangle\,,\label{eq:4}
\end{align}
where $\boldsymbol{s}_{\boldsymbol{k}}=\boldsymbol{U}_{\boldsymbol{k}}^{T}\smash{\boldsymbol{p}_{\boldsymbol{k}}^{\parallel}}\boldsymbol{U}_{\boldsymbol{k}}$
is the transform to the diagonal boson representation of the polarization
vertex $\smash{\boldsymbol{p}_{\boldsymbol{k}}^{\parallel}}=(\boldsymbol{p}_{x,\boldsymbol{k}},\boldsymbol{p}_{y,\boldsymbol{k}})\cdot\boldsymbol{e}_{E}$,
projected onto the in-plane electric field. The time ordering allows
for normal
\begin{equation}
-\langle T_{\tau}(D_{\boldsymbol{k}\mu}(\tau)D_{\boldsymbol{k}'\nu}^{\dagger})\rangle=\delta_{\mu\nu}\delta_{\boldsymbol{k}\boldsymbol{k}'}G_{\mu}(\boldsymbol{k},\tau)\,,\label{eq:5}
\end{equation}
as well as anomalous contractions
\begin{align}
-\langle T_{\tau}(D_{\boldsymbol{k}\mu}(\tau)D_{\boldsymbol{k}'\nu})\rangle= & -\langle T_{\tau}(D_{\boldsymbol{k}\mu}(\tau)D_{-\boldsymbol{k}'\bar{\nu}}^{\dagger})\rangle\nonumber \\
= & \,\delta_{\mu\bar{\nu}}\delta_{\boldsymbol{k},-\boldsymbol{k}'}G_{\mu}(\boldsymbol{k},\tau)\,,\label{eq:6}
\end{align}
where $\mu=1,2$, and we define $\bar{\nu}=(\nu+1$ mod $2)$. Finally,
the adjoint anomalous contractions satisfy $-\langle T_{\tau}(D_{\boldsymbol{k}\mu}^{\dagger}(\tau)\allowbreak D_{\boldsymbol{k}'\nu}^{\dagger})\rangle=\allowbreak-\langle T_{\tau}(D_{-\boldsymbol{k}\bar{\mu}}(\tau)\allowbreak D_{\boldsymbol{k}'\nu}^{\dagger})\rangle=\allowbreak\delta_{\mu\bar{\nu}}\delta_{\boldsymbol{k},-\boldsymbol{k}'}\allowbreak G_{\mu}(\boldsymbol{k},\tau)$,
using $\delta_{\bar{\mu}\nu}=\delta_{\mu\bar{\nu}}$.

\begin{figure}[tb]
\centering{}\includegraphics[width=0.7\columnwidth]{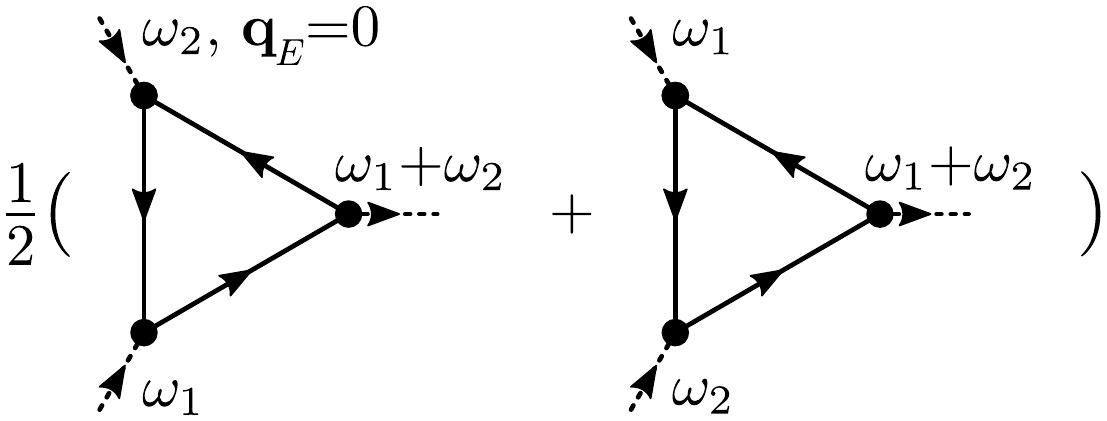}\caption{Diagrams for the 2D NLRF $\chi_{2}(\omega_{1},\omega_{2})$ at order
$O(2)$ in $E_{ac}(t)$. The solid lines carry an index $\mu=1,2$,
referring to $G_{1,2}(\boldsymbol{k},i\omega_{n})$ of Eq. (\ref{eq:8}),
the dots refer to the $2\times2$ polarization operator matrices $s_{\boldsymbol{k}\mu\nu}$.\label{fig:digs}}
\end{figure}

Any diagram at $O(n)$ is a closed loop of $n{+}1$ bilinear vertices,
linked by a normal or anomalous contraction. Since the anomalous Green's
functions are related to the normal ones by a mere shift of either
the right or the left matrix indices $\mu\rightarrow\bar{\mu}$ and
a flip $k'\rightarrow-k'$ of the sign of the momentum, the necessary
contractions can be reduced to only normal ones, by relabeling the
summation indices appropriately and by using a symmetrized vertex
$s_{\boldsymbol{k}\mu\nu}\rightarrow t_{\boldsymbol{k}\mu\nu}=(s_{\boldsymbol{k}\mu\nu}+s_{-\boldsymbol{k}\bar{\nu}\bar{\mu}})/2$
. With some algebra, one can show that $s_{-\boldsymbol{k}\bar{\nu}\bar{\mu}}=s_{\boldsymbol{k}\mu\nu}$,
simplifiying $t_{\boldsymbol{k}\mu\nu}=s_{\boldsymbol{k}\mu\nu}$.
Therefore, all diagrams can be generated using the matrix vertices
$s_{\boldsymbol{k}\mu\nu}$ and a 'single-arrowed' normal Green's
function $\boldsymbol{G}(\boldsymbol{k},i\omega_{n})$ of diagonal
matrix shape, with entries
\begin{align}
\boldsymbol{G}(\boldsymbol{k},i\omega_{n})= & \delta_{\mu\nu}G_{\mu}(\boldsymbol{k},i\omega_{n})\label{eq:8}\\
G_{1}(\boldsymbol{k},i\omega_{n})= & \frac{1}{i\omega_{n}-\epsilon_{\boldsymbol{k}}}\nonumber \\
G_{2}(\boldsymbol{k},i\omega_{n})= & G_{1}(-\boldsymbol{k},-i\omega_{n})\,.\nonumber 
\end{align}
The diagrams for the $O(2)$ 2D NRF are shown in Fig. (\ref{fig:digs}).
Since each single-arrowed line symbolizes two Green's functions, i.e.,
$G_{1(2)}(\boldsymbol{k},i\omega_{n})$, the diagram is a sum of $2^{3}=8$
expressions, each comprising one internal Matsubara frequency summation.
We obtain
\begin{align}
\chi_{2}(\omega_{1},\omega_{2})=\sum_{\boldsymbol{k}} & \left[\frac{8(1{+}2n_{\boldsymbol{k}})\epsilon_{\boldsymbol{k}}^{2}\,s_{\boldsymbol{k}11}\,|s_{\boldsymbol{k}12}|^{2}}{(z_{1}^{2}-4\epsilon_{\boldsymbol{k}}^{2})(z_{2}^{2}-4\epsilon_{\boldsymbol{k}}^{2})}\right.\nonumber \\
 & \times\left.\frac{(z_{1}^{2}{+}z_{1}z_{2}{+}z_{2}^{2}-12\epsilon_{\boldsymbol{k}}^{2})}{((z_{1}{+}z_{2})^{2}-4\epsilon_{\boldsymbol{k}}^{2})}\right]\,,\label{eq:chi2}
\end{align}
with the Bose function $n_{\boldsymbol{k}}=1/(\exp(\epsilon_{\boldsymbol{k}}/T)-1)$ and $z_{1,2}=\omega_{1,2}+i\eta$
complex frequencies, close to the real axis with $\omega_{1,2}\in\Re$
and $0<\eta\ll$1. For the model at hand, and since $\boldsymbol{U}_{\boldsymbol{k}}$
and $\boldsymbol{p}_{x(y),\boldsymbol{k}}$ are both real, we have
$|s_{\boldsymbol{k}12}|^{2}=s_{\boldsymbol{k}12}^{2}$. Eq. (\ref{eq:chi2})
is a main result of this work and it completes the evaluation of $\chi_{2}(\omega_{1},\omega_{2})$.

We mention that for the present bare LSWT approach, and not considering
magnon interactions, $\chi_{2}(\omega_{1},\omega_{2})$ can also be
evaluated calculating commutators. This is listed in App. \ref{sec:comm}.
Including $1/S$ corrections however, a diagrammatic approach is superior.

\section{Discussion\label{sec:Discussion}}

We begin our discussion by recalling that in principle, and due to
the Mermin-Wagner-Hohenberg theorem, breaking a continuous symmetry
in $D\leq2$ at $T\neq0$ is not possible. In practice, and for $D=2$,
this tends to manifest itself by weak finite-$T$ logarithmic singularities
arising from integration over poles $\sim1/\epsilon_{\boldsymbol{k}}^{2}$
from Goldstone zeros. To cure this, and instead of either allowing
for only $T=0$ or including a third dimension in any calculation,
a pragmatic remedy is to enforce a phenomenological cutoff $\epsilon_{c}$,
with $\epsilon_{\boldsymbol{k}}>\epsilon_{c}>0$. The energy gap $\epsilon_{c}$
mimics a finite correlation length, i.e., only quasi-long-range-order.
Implementing such a cutoff in the LSWT for the ICS is straightforward.
I.e., we replace $\mathcal{A}_{\boldsymbol{k}}\rightarrow\mathcal{A}_{\boldsymbol{k}}+d\,S/8j$
in Eq. (\ref{Ak}) with a free parameter $d$. This modifies $\epsilon_{\boldsymbol{k}}$
from Eq. (\ref{ek}) such that the expression $4+32j^{2}$ on the
first line is replaced by $d+4+32j^{2}$. This opens a gap of $O(\sqrt{d})$
at $\boldsymbol{k}=\boldsymbol{0}$ and $\boldsymbol{q}$. The formal
expressions for the Bogoliubov transformation remains unchanged, only
with $\mathcal{A}_{\boldsymbol{k}}$ and $\epsilon_{\boldsymbol{k}}$
modified as just described. 

Before continuing, we clarify that the preceding discussion is indeed
relevant also for $\chi_{2}(\omega_{1},\omega_{2})$, irrespective
of any resonance conditions for $\omega_{1,2}$. I.e., we count the
powers of $1/\epsilon_{\boldsymbol{k}}$ from the numerator of Eq.
(\ref{eq:chi2}). The vertex $s_{\boldsymbol{k}\mu\nu}$ comprises
two Bogoliubov transforms Eq. (\ref{eq:bogo2}), implying one factor
of $1/\epsilon_{\boldsymbol{k}}$ per vertex. Moreover, at any \emph{finite}
temperature and for $\epsilon_{\boldsymbol{k}}\ll$T, the Bose function
provides for an additional factor of $1/\epsilon_{\boldsymbol{k}}$.
Therefore, in the vicinity of the Goldstone zeros, where $\epsilon_{\boldsymbol{k}}\sim k$,
with $k=|\boldsymbol{k}|$, the numerator of Eq. (\ref{eq:chi2})
scales $\sim1/k$ for $\epsilon_{\boldsymbol{k}}\gg T$ and $\sim1/k^{2}$
for $\epsilon_{\boldsymbol{k}}\ll T$. I.e., at $T\neq$0 and independent
of $\omega_{1,2}$, the 2D momentum integration for $\chi_{2}(\omega_{1},\omega_{2})$
would be log-singular for $\epsilon_{c}=0$.

Next, we compare our results to recent analysis of 2D NRFs in Kitaev
magnets \citep{Brenig2024}. While these are fractionalized spin systems
with Majorana fermion and vison elementary excitations, it is reassuring
to realize that the 2D NRF and that of the present work are expressions
of identical form, except for the difference in statistics. Therefore,
and similar to ref. \citep{Brenig2024}, $\chi_{2}(\omega_{1},\omega_{2})$
along the so-called rectification or galvanoelectric (GEE) line, $\omega\equiv\omega_{1}=-\omega_{2}$,
is anomalously singular. In fact, and asymptotically for $\eta\ll1$,
Eq. (\ref{eq:chi2}) on the GEE line can be cast into
\begin{align}
\lefteqn{\chi_{2}(\omega{,}{-}\omega)\simeq\frac{\pi}{\eta}\sum_{\boldsymbol{k}}\left[(1{+}2n_{\boldsymbol{k}})\,s_{11\boldsymbol{k}}\times\right.}\nonumber \\
 & \hphantom{aaaaaaaaaa}\left.|s_{12\boldsymbol{k}}|^{2}(\delta(\omega+2\epsilon_{\boldsymbol{k}})+\delta(\omega-2\epsilon_{\boldsymbol{k}}))\right].\label{eq:c2gee2}
\end{align}
This function is purely real, consisting of a form factor weighted
two-magnon density of states, multiplied by a prefactor $\propto1/\eta$.
The latter implies a globally singular NRF on the GEE line for $\eta\rightarrow0^{+}$.
This is consistent with similar behavior reported in the context of
very different physical questions and for other systems \citep{Parker2019,Sipe2000,Fei2020,Ishizuka2022,Raj2024}.
We adopt the rational of the latter works and replace the causal broadening
$\eta$ with a \emph{physical scattering rate} $\Gamma$ in order
to render $\chi_{2}(\omega_{1},\omega_{2})$ finite along the GEE
line. Corrections beyond LSWT, e.g., magnon self-energies may provide
for finite $\Gamma$, but also extrinsic scattering from, e.g., lattice
degrees of freedom. In the following, we keep $\Gamma$ a free, momentum
and temperature independent parameter.

\begin{figure}[tb]
\centering{}\includegraphics[width=0.7\columnwidth]{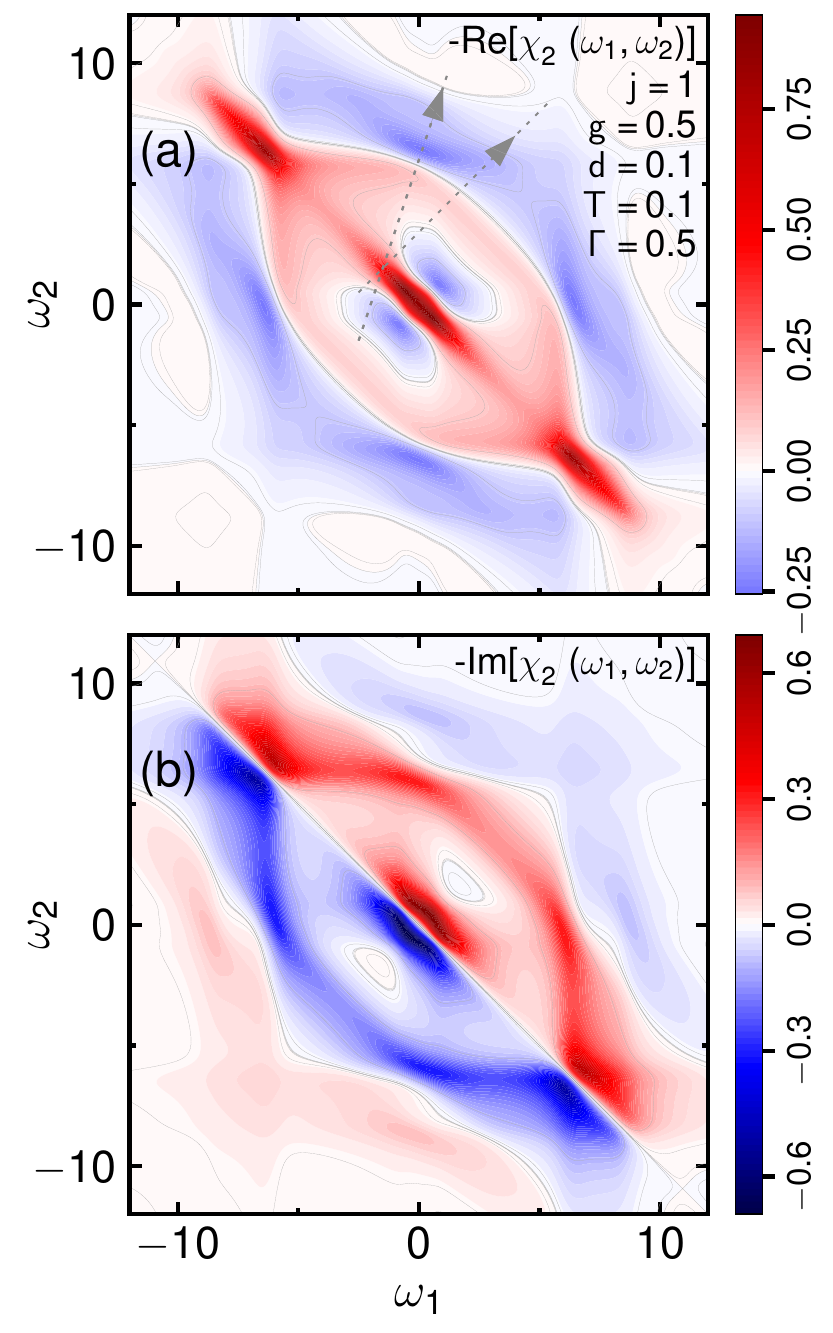}\caption{Contours of (a) the real and (b) imaginary part of the 2D NRF $\chi_{2}(\omega_{1},\omega_{2})$
at intermediate damping. Contour lines added at $\pm(.1,.01,.001,0)$
for better visibility of low amplitude structures. $S\equiv1$ to
ease notation. Thin dashed lines with arrow refer to cuts in Fig.
\ref{fig:2dcut}.\label{fig:NRF1}}
\end{figure}

Fig. \ref{fig:NRF1} displays contours of $\chi_{2}(\omega_{1},\omega_{2})$.We
set $\boldsymbol{E}_{ac}(t)=\boldsymbol{e}_{x}E_{ac}(t)$ hereafter,
i.e., only $P_{x}$ is used. For the parameters chosen, $\Gamma/\max(\epsilon_{\boldsymbol{k}})\lesssim.1$.
While this seems a realistic and not too small damping rate, it has
been set primarily such as to keep the singular behavior along the
GEE line on the same scale as other aspects of the plot. First, the
overall dominant characteristic of $\chi_{2}(\omega_{1},\omega_{2})$
is its strong antidiagonal amplitude related to Eq. (\ref{eq:c2gee2}).
Obviously, this feature is not confined strictly to $\omega_{2}=-\omega_{1}$.
In fact, and as to be expected from the prefactor of $1/\Gamma$ in
Eq. (\ref{eq:c2gee2}), smoothly continuing $\chi_{2}(\omega_{1},\omega_{2})$
into the 2D frequency plane, perpendicularly off the GEE line, the
prefactor can be viewed asymptotically as the limiting form of some
Lorentzian $\sim i/(\omega_{\perp}+i\Gamma)$ with a frequency variable
$\omega_{\perp}$, perpendicular to the GEE line. This captures both,
the real part $\sim1/\Gamma$, as well as the imaginary part being
strictly zero with a sign change across the GEE line which is clearly
visible in Fig. \ref{fig:NRF1}. Therefore, and as another main result
of this work, quasiparticle damping rates can be read of from 2D NRFs
perpendicular to the GEE line. We emphasize that while we use a constant
rate $\Gamma$, it is straightforward to extend the discussion of
Eq. (\ref{eq:c2gee2}) to assume a momentum dependence $\Gamma_{\boldsymbol{k}}$.
In that case, the Lorentzian's FWHM in $\omega_{\perp}$ perpendicular
to the antidiagonal at various locations $\omega=\pm\epsilon_{\boldsymbol{k}}$
represents a momentum-resolved analysis.

Along the GEE line, Fig. \ref{fig:NRF1} allows to read off the band-width
of $2\epsilon_{\boldsymbol{k}}$. This is $\approx$10.7 for the parameters
chosen and indeed, along the antidiagonal the real part of the NRF
approaches zero in that region. The intensity variations visible along
the GEE line are a combination of matrix element- and dispersion-effects.
Finally, Fig. \ref{fig:NRF1} shows a strong ring-like pattern involving
finite $\omega_{1,2}$. This is related to energies from the region
of the BZ corners $\boldsymbol{k}\sim(\pi,\pi)$, see Fig. \ref{fig:dispplt}.
Since their magnitude varies only weakly for sufficiently large $j$,
e.g., $\epsilon_{\pi,\pi}=(1/j-4)$, this ring pattern remains located
roughly within the same frequency range for all $j\gtrsim1$. 

\begin{figure}[tb]
\centering{}\includegraphics[width=0.7\columnwidth]{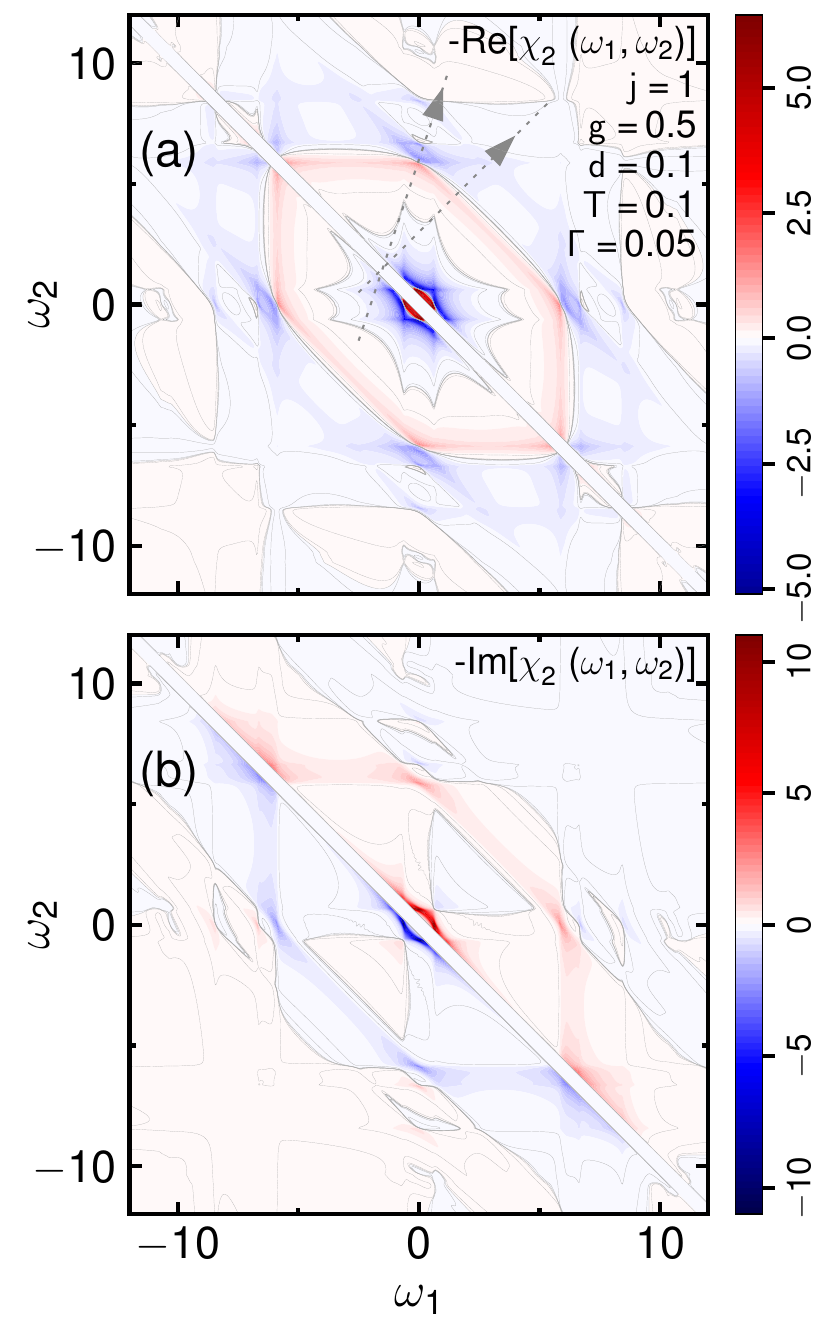}\caption{Contours of (a) the real and (b) imaginary part of the 2D NRF $\chi_{2}(\omega_{1},\omega_{2})$
at small damping. Contour lines added at $\pm(.1,.01,.001,0)$ for
better visibility of low amplitude structures. $S\equiv1$ to ease
notation. Thin dashed lines with arrow refer to cuts in Fig. \ref{fig:2dcut}.
\label{fig:NRF2}}
\end{figure}

In order to uncover also finer details and for completeness, Fig.
\ref{fig:NRF2} displays the 2D NRF for a damping $\Gamma$, which
is one order of magnitude smaller than in Fig. \ref{fig:NRF1}. This
implies a huge and very sharp GEE response, suppressing the visibility
of all smaller intensities in the figure. For visualization purposes,
we therefore mask the NRF in a narrow strip along the GEE line to
zero, which moves the NRF's fine structure up front. First, at this
damping, the figure demonstrates more clearly that the boundaries
of the larger ring-like feature are set by the density of states at
the energies from the region of the BZ corners $\boldsymbol{k}\sim(\pi,\pi)$.
In fact, for the parameters chosen, $2\epsilon_{\pi,\pi}\approx6$,
which from the energy denominator in Eq. (\ref{eq:chi2}) leads to
six exceptional frequency locations of $(\omega_{1},\omega_{2})\approx(\pm6,0),(0,\pm6),(\pm6,\mp6)$,
harboring sign changes of $\mathrm{Re}[\chi_{2}(\omega_{1},\omega_{2})]$,
which can be observed in Fig. \ref{fig:NRF2} to set the ring feature.
An additional location of a far weaker NRF amplitude with a sign change
of $\mathrm{Re}[\chi_{2}(\omega_{1},\omega_{2})]$ is visible at $(\omega_{1},\omega_{2})\approx(\pm6,\pm6)$.
Second, the most intense feature is a small ring-like structure confined
by exceptional energies $\omega_{1,2}\sim O(0.6)$. Part of it is
masked to zero on the GEE line. These energies result from the correlation
gap, primarily from $\boldsymbol{k}\sim\boldsymbol{q}$, where $2(\epsilon_{\boldsymbol{q}})\approx0.6$.
Finally, no obvious exceptional frequencies can be identified which
relate to the maximum of the LSWT dispersion, i.e., $2\max(\epsilon_{\boldsymbol{k}})\approx10.7$,
which is clearly within axes limits of $|\omega_{1,2}|\leq12$ in
Fig. \ref{fig:NRF2}.

\begin{figure}[tb]
\centering{}\includegraphics[width=0.8\columnwidth]{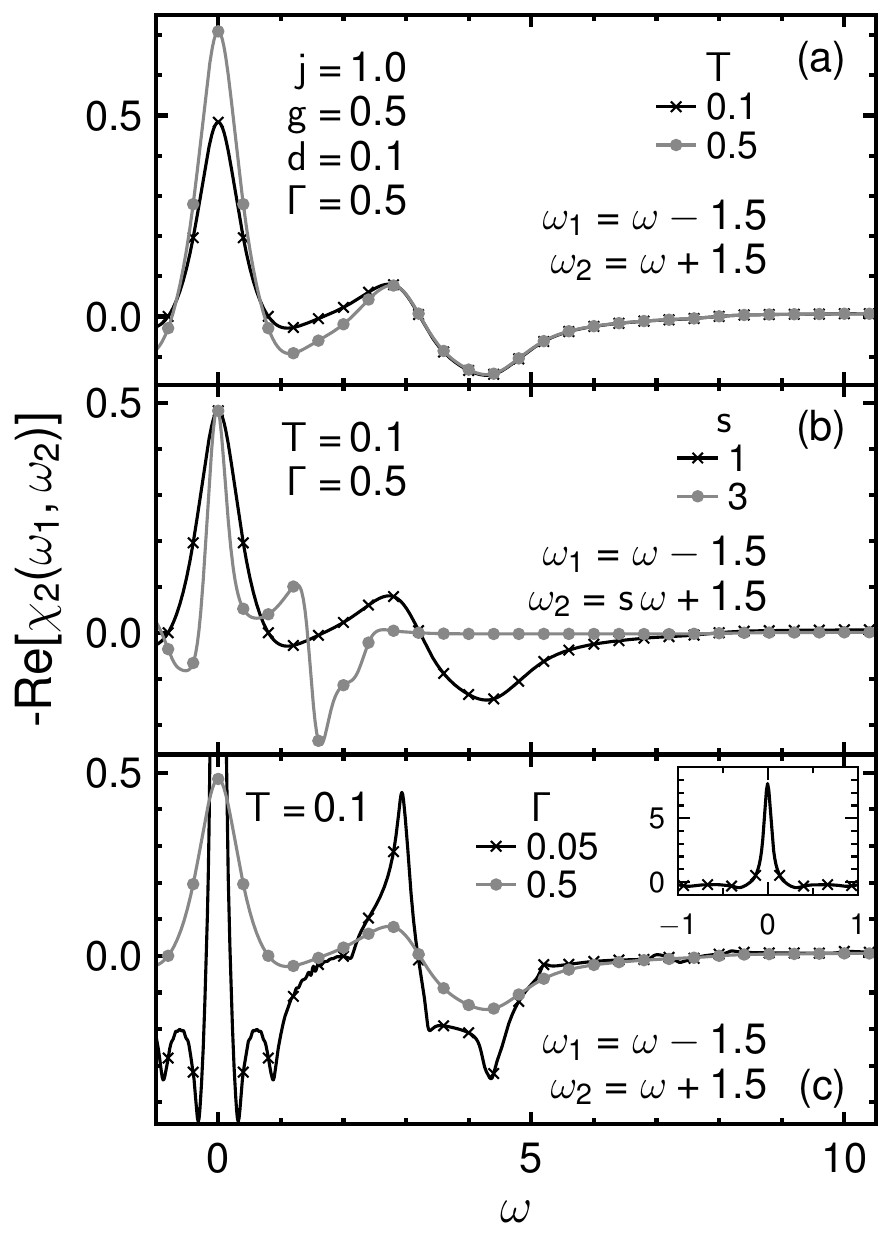}\caption{Cuts of the real part of the 2D NRF $\chi_{2}(\omega_{1},\omega_{2})$
along the dashed directions in Figs. \ref{fig:NRF1},\ref{fig:NRF2}
versus two (a) temperatures, (b) directions, and (c) damping rates.
$S\equiv1$ to ease notation.\label{fig:2dcut}}
\end{figure}

In Fig. \ref{fig:2dcut}, various cuts are depicted through the real
part of 2D NRF along the paths shown with dashed gray arrows in Figs.
\ref{fig:NRF1}, \ref{fig:NRF2}. To begin, these paths have been
shifted off from the region $\omega_{1,2}\ll1$, not to be dominated
by the correlation gap at small $\omega$. Next, panel (a) shows a
typical behavior perpendicular to the GEE line at the damping $\Gamma$
used also in Fig. \ref{fig:NRF1}. Very clearly the panel displays
a quasi-Lorentzian with a FWHM $\sim\Gamma$ at $\omega=0$. This
has been anticipated in the discussion of Eq. (\ref{eq:c2gee2}),
and Fig. \ref{fig:NRF1} and may potentially allow to extract $\Gamma$
from experiment. Moreover, the panel shows the impact of the Bose
statistics of the magnons. I.e., increasing temperature, the low-energy
amplitudes increase, however, the high-energy structure remains unchanged.
Panel (b) shows the dependence of $\mathrm{Re}[\chi_{2}(\omega_{1},\omega_{2})]$
on different cut directions, and in particular the abrupt sign changes
at exceptional frequencies in the context of the ring structures,
discussed for Figs. \ref{fig:NRF1}, \ref{fig:NRF2}. I.e., for the
path with $(\omega_{1},\omega_{2})=(\omega{-}1.5,3\omega{+}1.5)$,
the exceptional point $(\omega_{1},\omega_{2})\approx(0,6)$ is reached
at $\omega\approx1.5$, where indeed Fig. \ref{fig:2dcut}(b) displays
a rapid sign change of $\mathrm{Re}[\chi_{2}(\omega_{1},\omega_{2})]$.
The latter can be increased by decreasing $\Gamma$. Finally, panel
(c) is primarily intended to show the scaling of the NRF amplitude
with $\Gamma$ close to, but perpendicular to the GEE line. For that
purpose, the same values of $\Gamma$ as used in Figs. \ref{fig:NRF1},
\ref{fig:NRF2} are chosen. For $\Gamma=0.05$, we cut off $\mathrm{Re}[\chi_{2}(\omega_{1},\omega_{2})]$
at low-$\omega$ and redisplay it in the inset. Cum grano salis, the
ratio of the peak intensities of $\mathrm{Re}[\chi_{2}(\omega_{1},\omega_{2})]$
on the GEE line for $\Gamma_{\times}/\Gamma_{\bullet}=0.1$ is approximately
$10$, in agreement with Eq. (\ref{eq:c2gee2}). The same can be read
off for the ratio of the FWHMs. Apart from that, Fig. \ref{fig:2dcut}(c)
obviously shows that $\Gamma=0.5$ already leads to a significant
smearing of fine-structure of the NRF.

\section{Summary\label{sec:Summary}}

To recap, we have argued that spiral magnets are a promising testbed
for coherent nonlinear optical spectroscopy. In such magnets, spin-current
coupling can provide for a dominant effective dipole moment which
allows for a minimal coupling to electric fields and generates a nonlinear
response already at second order in the driving field. We have detailed
the consequences of this argument for the spiral phase of the square
lattice $J_{1}$-$J_{3}$ Heisenberg antiferromagnet, using linear
spin wave theory and evaluating the second order response function.
Apart from a rich landscape in the 2D frequency plane, we have found
the prime feature of this response function to be a dominant antidiagonal
structure which allows to read off quasiparticle lifetimes. This may
be of interest for experiments. Obvious extensions of our study could
include magnon-selfenergy corrections, higher-order response functions,
and other magnets with spiral order or correlations.
\begin{acknowledgments}
Fruitful discussions with Anna Keselman, Johannes Knolle, Peter Orth,
and Natalia Perkins are gratefully acknowledged. We thank A. Schwenke
for critical reading of the manuscript. Research of W.B. was supported
in part by the DFG through Project A02 of SFB 1143 (project-id 247310070)
and by NSF grant PHY-2210452 to the Aspen Center for Physics (ACP).
W.B. acknowledges kind hospitality of the PSM, Dresden.
\end{acknowledgments}

\appendix

\section{Commutator approach\label{sec:comm}}

It has become customary in part of the literature to discuss the 2D
Fourier transform of the retarded response function $\tilde{\chi}_{2}(t,t_{1},t_{2})=i^{2}\Theta(t-t_{1})\Theta(t_{1}-t_{2})\allowbreak\langle[[P^{\parallel}(t),\allowbreak P^{\parallel}(t_{1})],P^{\parallel}(t_{2})]\rangle$
directly, discarding intrinsic permutation symmetry. As long as $P^{\parallel}(t)$
can be expressed in terms of known eigenstates and energies, e.g.,
for free quasiparticles as in the present case of non-interacting
LSWT, calculating the commutators is an alternative to calculating
the diagrams of Fig \ref{fig:digs}. However, in case of non-diagonal
quasiparticle interactions, approximate approaches to evaluating the
commutator correlation function are conceptually unclear, rendering
the diagrammatic approach superior in principle at least.

\begin{figure}[!tb]
\centering{}\vskip .2cm\includegraphics[width=0.99\columnwidth]{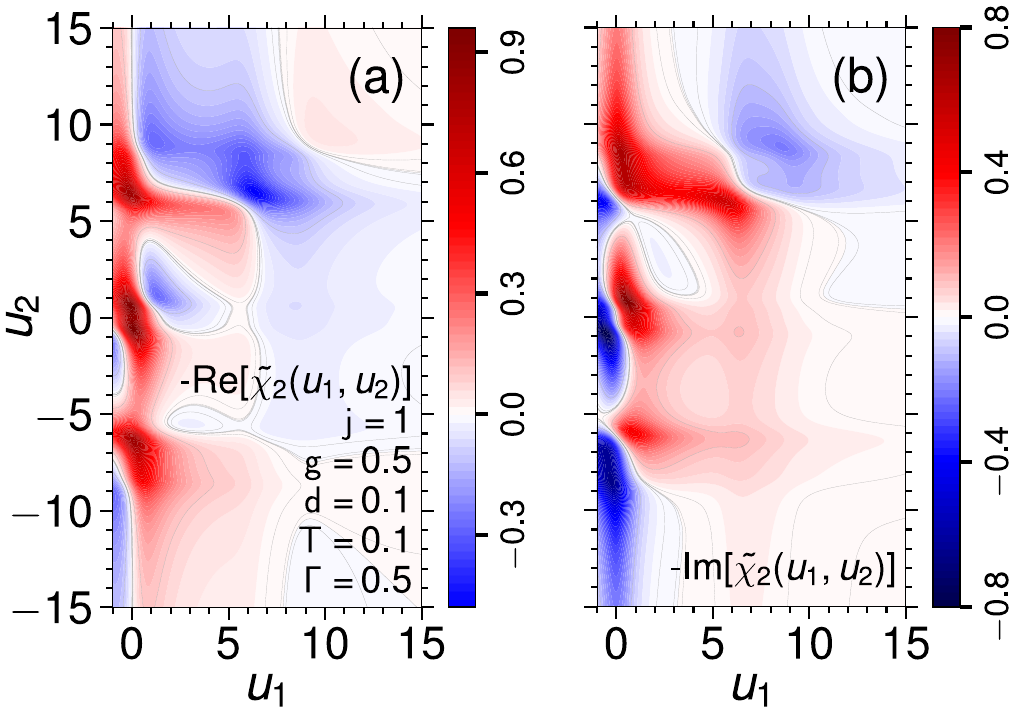}\caption{Contours of (a) the real and (b) imaginary part of the commutator
correlation function $\tilde{\chi}_{2}(u_{1},u_{2})$ at intermediate
damping. Contour lines added at $\pm(.1,.01,.001,0)$ for better visibility
of low amplitude structures. $S\equiv1$ to ease notation.\label{fig:comcont}}
\end{figure}

For completeness, we list the result for $\tilde{\chi}_{2}(\omega_{1},\omega_{2})$
that we find from the commutator expression using the time dependent
$P^{\parallel}(t)$ of the diagonal boson representation of the LSWT
\begin{align}
\tilde{\chi}_{2}(\omega_{1},\omega_{2})=\sum_{\boldsymbol{k}} & \left[\frac{16(1{+}2n_{\boldsymbol{k}})\epsilon_{\boldsymbol{k}}^{2}\,s_{\boldsymbol{k}11}\,|s_{\boldsymbol{k}12}|^{2}}{(z_{1}+z_{2})(z_{2}^{2}-4\epsilon_{\boldsymbol{k}}^{2})}\right.\nonumber \\
 & \times\left.\frac{(z_{1}{+}2z_{2})}{((z_{1}{+}z_{2})^{2}-4\epsilon_{\boldsymbol{k}}^{2})}\right]\,.\label{eq:chi2comm}
\end{align}
Indeed and most important, $(\tilde{\chi}_{2}(\omega_{1},\omega_{2})+\tilde{\chi}_{2}(\omega_{2},\allowbreak\omega_{1}))/2=\chi_{2}(\omega_{1},\allowbreak\omega_{2})$,
identical to Eq. (\ref{eq:chi2}) is satisfied.

Apart from considering $\tilde{\chi}_{2}$ instead of $\chi_{2}$,
it has also become customary to Fourier-transform the latter with
respect to a different set of time coordinates, i.e., $t^{\prime}+\tau=t-t_{2}$
and $\tau=t_{1}-t_{2}$. This is motivated by a two $\delta$-function
pulse sequence, separated by time $\tau$ and followed by time a $t^{\prime}$
up to measurement. This changes the frequencies variables to $u_{1}=\omega_{1}+\omega_{2}$
and $u_{2}=\omega_{2}$. While we feel that contours of $\tilde{\chi}_{2}(u_{1},u_{2})$
are less intuitive to interpret, we also display them in Fig. \ref{fig:comcont},
using the parameters of Fig. \ref{fig:NRF1}. Typically, only the
first and fourth quadrant in the $u_{1}$,$u_{2}$-plane are displayed.
We follow this habit, yet, keeping a small portion of the negative
$u_{1}$-axis. The latter is in order to clarify that also in $\tilde{\chi}_{2}(u_{1,}u_{2})$
a strong GEE intensity is observed, however, located along the $u_{2}$-axis,
at $u_{1}=0$. Other central features can be ``rediscovered'' by
comparing Figs. \ref{fig:comcont} and \ref{fig:NRF1} side-by-side.


\begin{thebibliography}{99}
\bibitem{ArbS}Nonetheless, our following calculations will be performed
for arbitrary spin.
\end{thebibliography}
\end{document}